\documentclass[epj,twocolumn]{webofc}
\usepackage[varg]{txfonts}   % Web of Conferences font
\usepackage{graphicx}
\wocname{epj}
\woctitle{Seismology of the Sun and the Distant Stars 2016}
\begin{document}
%***********************************************************************
\title{Asteroseismic inversions in the Kepler era: application to the Kepler Legacy sample}

\author{\firstname{Gaël} \lastname{Buldgen}\inst{1,3}\fnsep\thanks{\email{gbuldgen@ulg.ac.be}} \and
        \firstname{Daniel} \lastname{Reese}\inst{2}\fnsep\thanks{\email{daniel.reese@obspm.fr}} \and
        \firstname{Marc-Antoine} \lastname{Dupret}\inst{3}\fnsep\thanks{\email{ma.dupret@ulg.ac.be}}
}

\institute{Institut d’Astrophysique et Géophysique de l’Université de Liège, Allée du 6 août 17, 4000 Liège, Belgium \and LESIA, Observatoire de Paris, PSL Research University, CNRS, Sorbonne Universités, UPMC Univ. Paris 06, Univ. Paris Diderot, Sorbonne Paris Cité, 5 place Jules Janssen, 92195 Meudon Cedex, France}

%-----------------------------------------------------------------------
\abstract{In the past few years, the CoRoT and Kepler missions have carried out what is now called the space photometry revolution. This revolution is still ongoing thanks to K$2$ and will be continued by the Tess and Plato$2.0$ missions. However, the photometry revolution must also be followed by progress in stellar modelling, in order to lead to more precise and accurate determinations of fundamental stellar parameters such as masses, radii and ages. In this context, the long-lasting problems related to mixing processes in stellar interior is the main obstacle to further improvements of stellar modelling. In this contribution, we will apply structural asteroseismic inversion techniques to targets from the Kepler Legacy sample and analyse how these can help us constrain the fundamental parameters and mixing processes in these stars. Our approach is based on previous studies using the SOLA inversion technique \cite{Pijpers} to determine integrated quantities such as the mean density \cite{Reese}, the acoustic radius, and core conditions indicators \citep{Buldgentu}, and has already been successfully applied to the $16$Cyg binary system \cite{BuldgenA}. We will show how this technique can be applied to the Kepler Legacy sample and how new indicators can help us to further constrain the chemical composition profiles of stars as well as provide stringent constraints on stellar ages.}
\maketitle
%
%-----------------------------------------------------------------------
\section{Introduction}
\label{intro}
In the last decade, the space photometry missions CoRoT and Kepler have provided us with a wealth of precise seismic data for a large number of stars. This revolution is now allowing us to carry out precise seismic modelling for these stars and to test the physical inputs of our stellar models. As a by-product of this modelling, asteroseismology also delivers precise and accurate fundamental parameters, which are essential to other fields such as galactic archeology or exoplanetology.

Amongst the observed targets, the Kepler Legacy sample contains the dwarfs that have the best frequency sets within the Kepler data (See \cite{Lund} for the presentation of the dataset and \cite{Silva} for forward modeling results of the Kepler Legacy sample). As such, they can be seen as benchmark stars to test our inputs of stellar models and the efficiency of seismic modelling. Due to their high number of frequencies and the precision of these determinations, they are also the ideal targets with which to carry out structural seismic inversions. In this paper, we will present forward modelling and inversion results for two targets of the Legacy sample: namely Doris (KIC8006161) and Saxo (KIC6603624). In Sect. \ref{sec-1}, we present the targets and the forward modelling results for each one of them. In Sect. \ref{sec-2}, we present the inversion procedure we used to further constrain the models of both stars and present the results of this process in Sect. \ref{sec-3}. The implications of these results are then presented in Sect. \ref{sec-4}. 

\section{Targets selection and forward modelling}
\label{sec-1}

The targets were selected by the number and precision of their observed frequencies as well as their estimated mass from a crude grid-based modelling approach. We specifically selected stars with masses close to $1M_{\odot}$. Firstly, stars closer to the sun are easier to model, and we know from helioseismology and from our theoretical test cases where it was possible to obtain a sufficiently good modelling that linear inversion techniques can be applied to constrain their structure (see \cite{Basu09} for an application to the solar case). This is potentially not the case for more massive stars for which the boundary of the convective core may not be reproduced accurately enough and thus the evolutionary stage of the target not very well constrained. In such cases, using a linear approximation may be questionable. The second reason is that we wanted to avoid convective cores in the reference models because we used specific indicators using radial derivatives to constrain the internal structure of the targets \cite{Buldgentu}. The thermodynamical quantities used in these indicators can have discontinuous derivatives which will appear in the target function of the SOLA method. Such targets are then impossible to fit with the continuous structural kernels built using the eigenfunctions of solar-like oscillation modes.

Before trying to carry out structural inversions, we obtained reference models for each target using a classical seismic forward modelling approach. This modelling was carried out using a Levenberg-Marquardt algorithm. The models were built using the Liège stellar evolution code (CLES, \cite{ScuflaireCLES}) and the oscillations were computed using the Liège oscillation code (LOSC, \cite{Scuflaire}). We used the Ceff equation of state \cite{CEFF}, the Opal opacities \cite{Iglesias} supplemented at low temperatures by the opacities from \cite{Alexander} and the nuclear reaction rates from the NACRE project \cite{Nacre} including the updated reaction rate for the $^{14}\mathrm{N}(p,\gamma)^{15}\mathrm{O}$ reaction from \cite{Formicola}. Convection was implemented using the classical, local mixing-length theory \cite{Bohm}. We used the individual small frequency separations, $d_{02}$ and $d_{13}$, the effective temperature, $T_{eff}$, and the metallicity, $\left[ Fe/H \right]$ as constraints for our structural models. Some fits were also performed using the frequency ratios $r_{01}$ and $r_{02}$ \cite{Roxburgh}and this lead to very similar results. No surface correction was applied to the frequencies since we used quantities that are not strongly dependant on surface effects. The free parameters of the Levenberg-Marquardt algorithm were the mass of the star, $M$, its age, its hydrogen mass fraction, $X$, its metallicity, $Z$, and the mixing-length parameter, $\alpha_{MLT}$. Both stars were modelled using the AGSS09 abundance tables \cite{Asplund9}, but tests were performed using the old GN93 abundances \cite{Grevesse} to see whether some comments could be made on the metallicity scale used to relate stellar metallicity to solar metallicity.

\subsection{Doris a.k.a KIC8006161}

\begin{figure}[h]
\centering
\includegraphics[width=\hsize,clip]{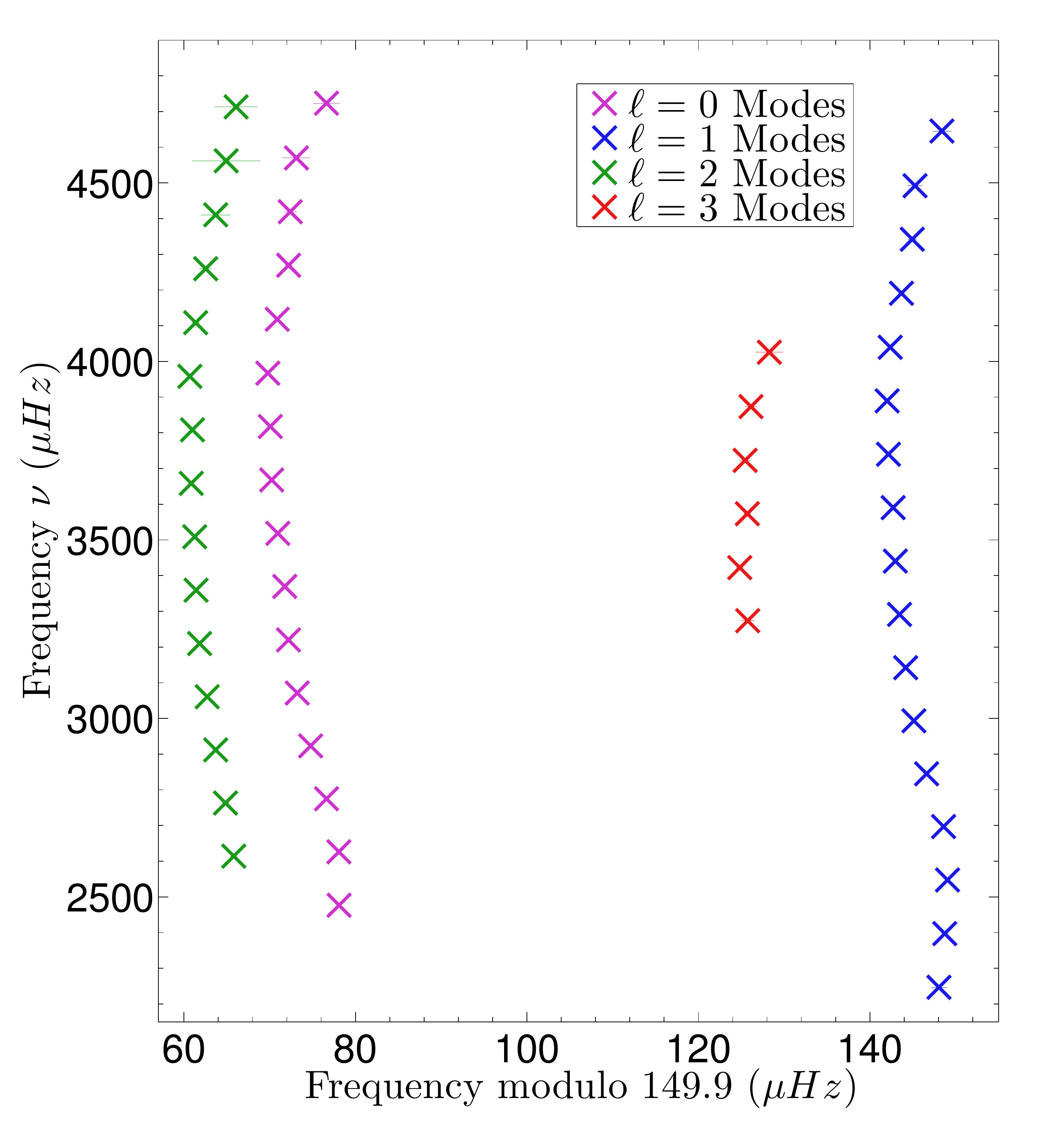}
\caption{Echelle diagram of Doris $(KIC8006161)$ illustrating the ridges of the $\ell=0,1,2,3$ modes.}
\label{figEchDor}       % Give a unique label
\end{figure}

In Fig. \ref{figEchDor}, we illustrate the echelle diagram of Doris. This star has $54$ observed individual frequencies with a mean $1 \sigma$ uncertainty of $0.49$ $\mu Hz$, an effective temperature of $5488 \pm 77K$ and a $\left[ Fe/H \right]=0.34 \pm 0.10$. The initial mass estimate was $0.96 M_{\odot}$. We illustrate in table \ref{tabRes} the results from the forward modelling process. These results are in agreement with the first estimate to within a reasonable accuracy. We note that most of the uncertainty stems from the well-known helium-mass degeneracy in seismic fitting and the intensity of diffusion. It seems also clear that extra-mixing processes acting inside the star would have an impact on these determinations. We only considered models with microscopic diffusion but without including turbulent diffusion. We note that it was possible to fit this target with both the $GN93$ abundances or the $AGSS09$ abundances within the same accuracy. Had we included extra-mixing, the scatter of the fundamental parameters would have been slightly larger.   
\begin{table}
\centering
\vspace{3pt} % if necessary!
\caption{Forward modelling results for Doris and Saxo}
\label{tabRes}       % Give a unique label
% For LaTeX tables you can use
\begin{tabular}{lll}
\hline
 & Doris & Saxo  \\\hline
Mass $(M_{\odot})$ & $0.91 - 1.02$ & $0.93-1.05$ \\
Age $(Gy)$ & $4.6-5.3$ & $7.6-8.7$ \\\hline
\end{tabular}
\end{table}

\subsection{Saxo a.k.a KIC6603624}

\begin{figure}[h]
\centering
\includegraphics[width=\hsize,clip]{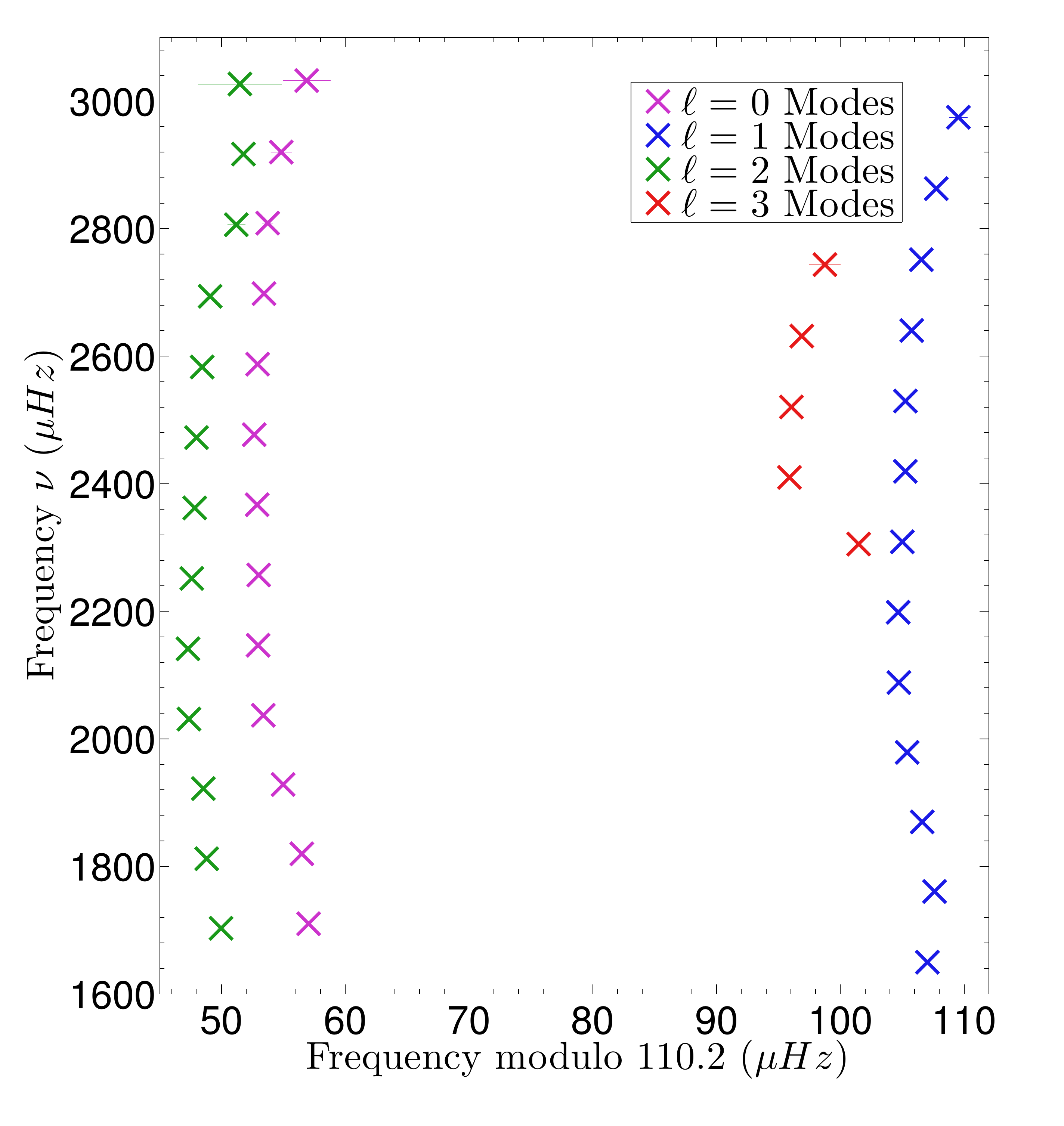}
\caption{Echelle diagram of Saxo $(KIC6603624)$ illustrating the ridges of the $\ell=0,1,2,3$ modes.}
\label{figEchSax}       % Give a unique label
\end{figure}

In Fig. \ref{figEchSax}, we illustrate the echelle diagram of Saxo. This target has $44$ individual frequencies with an average uncertainty of $0.336$ $\mu Hz$. Although the frequencies are accurately determined, we wish to point out a small deviation in the echelle diagram, the octupole mode of lowest frequency seems to be deviating from the ridge formed by the other octupole modes. This irregularity made this mode very difficult to fit and it seemed that it could be due to a miscalculation of the frequency value (Differences between different fitters have indeed been reported in \cite{RoxCyg}). It seemed unprobable that this difference could be physical so we eliminated this mode and used only $43$ frequencies to calculate the small frequency separations used in our fits. In addition to the seismic data, we used the same constraints as for Doris, namely the effective temperature, at a value of $5674 \pm 77$ and the $\left[ Fe/H \right]=0.28 \pm 0.10$. Our results, given in table \ref{tabRes} are in agreement with the initial mass estimate of $1.01 M_{\odot}$. Again, this target could be fitted using both the old GN$93$ and the new AGSS$09$ solar abundances. 
\section{Inversion Procedure}
\label{sec-2}
In this section we briefly present the inversion procedure we used for both targets. We followed the same methodology as in our study of the $16$Cyg binary system \cite{BuldgenA} and carried out inversions for both the mean density and a core condition indicator based on the derivative of the squared isothermal sound speed, denoted $t_{u}$. The mean density inversions have been described in \cite{Reese} and the $t_{u}$ inversions have been presented in \cite{Buldgentu}. The SOLA inversion technique \cite{Pijpers} was used to related the relative frequency differences to corrections of these integrated quantities, assuming the following relation
\begin{align}
\sum_{i}^{N}c_{i}\frac{\delta \nu_{i}}{\nu_{i}} \equiv \left(\frac{\delta A}{A}\right)_{inv}, \label{EqtuDefInv}
\end{align} 
where the $c_{i}$ are the inversion coefficients, found by minimizing the cost function of the SOLA method and $A$ is the integrated quantity whose correction is sought. Structural inversions in asteroseismology are carried out using a sample of reference models to analyze whether model dependency dominates the results or not. This is a consequence of the linearity hypothesis used to compute the structural corrections from seismic observations. To obtain the corrections of integrated quantities, we use the linear relations between frequency and structure from \cite{Gough}
\begin{align}
\frac{\delta \nu}{\nu}=\int_{0}^{R}K^{n,l}_{s_{A},s_{B}}\frac{\delta s_{A}}{s_{A}}dr + \int_{0}^{R}K^{n,l}_{s_{B},s_{A}}\frac{\delta s_{B}}{s_{B}}dr,
\end{align}
with $s_{A}$ and $s_{B}$ being structural quantities like the squared adiabatic sound speed, $c^{2}$, and the density $\rho$ as in classical helioseismic inversions. The $K^{n,l}$ functions are the structural kernels, which depend only on the reference model and its eigenfunctions. The $\delta$ notation is related to the differences between observed and reference quantities following the convention
\begin{align}
\frac{\delta x}{x}=\frac{x_{obs}-x_{ref}}{x_{ref}},
\end{align}
where $x$ can be a frequency or a structural quantity. The main problem in the context of asteroseismology is that the linearization hypothesis may not be valid. Thus, one can use multiple reference models to check for the robustness of the inference made with the inversion technique.

The definition of the $t_{u}$ indicator is
\begin{align}
t_{u}=\int_{0}^{R}f(r)\left(\frac{du}{dr}\right)^{2}dr \label{EqtuRef},
\end{align}
with $f(r)= r(r-R)^{2}\exp(-7r^{2})$, the weight function used for this inversion, $R$ the stellar radius, $r$ the radial coordinate associated with each layer inside the model, and $u$ the squared isothermal sound-speed defined as $u=\frac{P}{\rho}$. This quantity is very sensitive to changes in the deep layers of stars and can be used to assess the reliability of the representation of the stratification of the deep layers by the models. In fact, one can show that $u\approx \frac{T}{\mu}$ in the core regions and thus, this quantity will be very sensitive to the effects of mixing processes that will change the temperature, $T$, and mean molecular weight, $\mu$, gradients.

\begin{figure*}
\centering
\includegraphics[width=\hsize,clip]{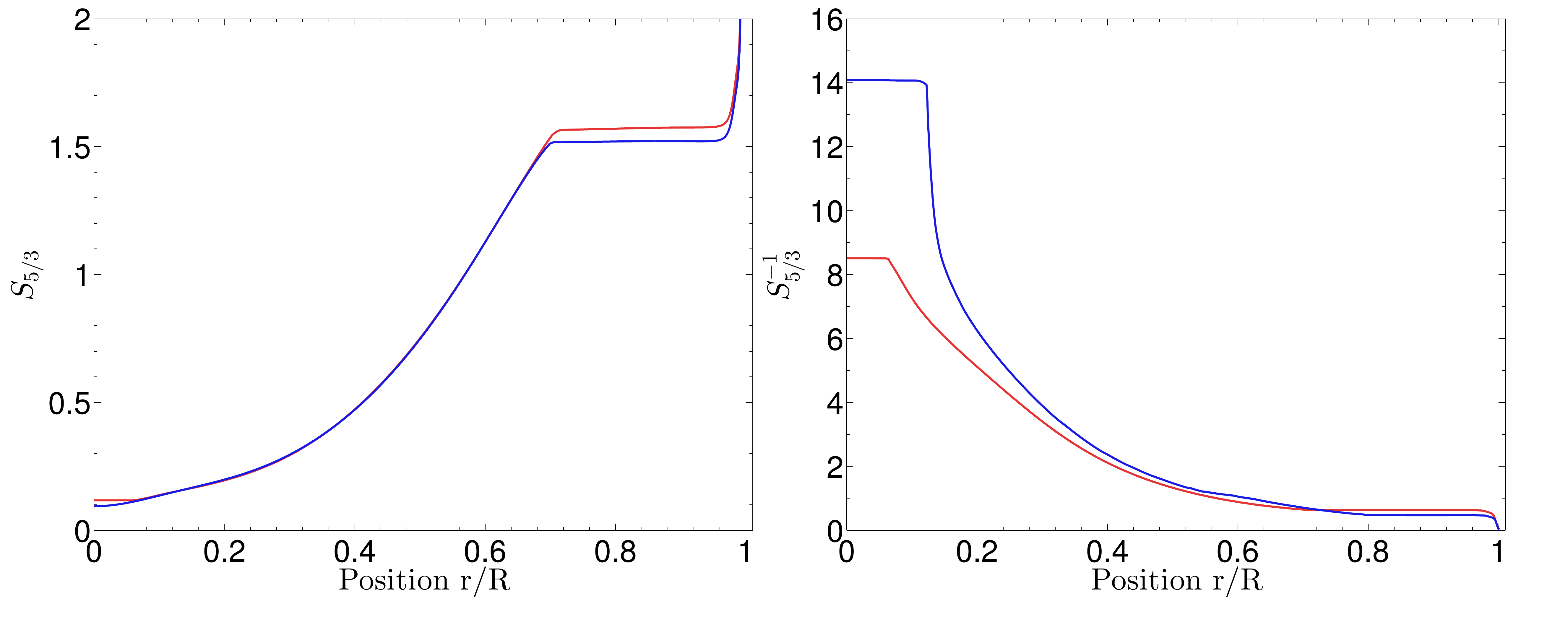}
\caption{Left panel: radial profile of $S_{5/3}$ for two stellar models of various mass and age, illustrating the plateau in the convective region. Right panel: radial profile of $S^{-1}_{5/3}$ for two stellar models, illustrating the plateau of the proxy in convective cores.}
\label{figSstruc}       % Give a unique label
\end{figure*}

Besides this indicator, we also carried out inversions using a new core condition indicator based on an entropy proxy, denoted $S_{5/3}=\frac{P}{\rho^{5/3}}$. We consider the inversion results of this indicator to be preliminary and a follow up with the definition and limitations of this inversion will be presented in a future paper. This proxy is particularly well suited for stars with a convective core since it will show a plateau in regions where convection dominate. The width of the plateau is linked to the extent of the convective regions and the height is sensitive to the transition with radiative regions, thus to the temperature and chemical composition gradient. In Fig \ref{figSstruc}, we show the behaviour of $S_{5/3}$ for the convective envelope in the left panel and of $S_{5/3}^{-1}$ in the right panel for a convective core.
\section{Inversion Results}
\label{sec-3}
\begin{figure*}[h]
\centering
\includegraphics[width=\hsize,clip]{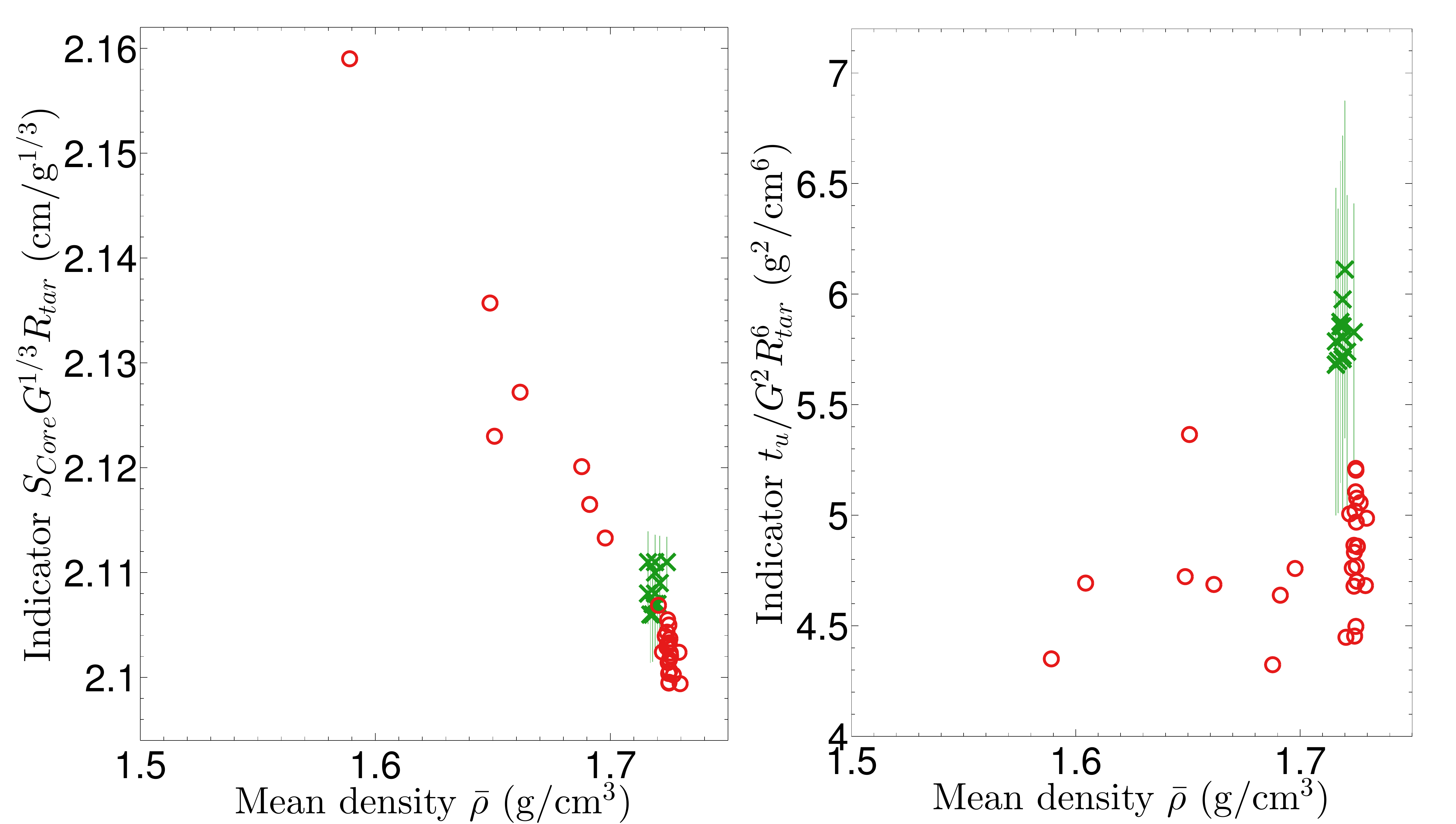}
\caption{Inversion results for Doris. Left panel: $\bar{\rho}-S_{Core}G^{1/3}R_{tar}$ plane where the red circles illustrate the position of the reference models and the green crosses show the inversion results with their error bars. Right panel: $\bar{\rho}-t_{u}/(G^{2}R^{6}_{tar})$ plane following the same notations as the left panel.}
\label{FigResDoris}       % Give a unique label
\end{figure*}

In Fig. \ref{FigResDoris}, we illustrate the inversion results for the $t_{u}$ indicator and an indicator based on $S_{5/3}^{-1}$, denoted $S_{Core}$. The green crosses are the inverted results while the red circles illustrate the reference models computed using the Levenberg-Marquardt algorithm. As can be seen, some models are in agreement with the inversion results while others are not. There is also a clear trend showing that models not fitting the inverted mean density, given here in abscissa, are incompatible with the other indicators. As already seen in \cite{BuldgenA}, the error bars on the $t_{u}$ inversion are quite large, but there is a clear trend that can be deduced and used to select a subsample of models from our initial forward modelling. The numerical value obtained for the inverted $t_{u}/G^{2}R_{Tar}^{6}$ value is around $5.8\pm 0.7 g^{2}/cm^{6}$, with $R_{Tar}$ the target photospheric radius and $G$ the graviational constant. The $S_{Core}$ indicator shows an inverted $S_{Core}GR_{Tar}^{1/3}$ value around $2.110 \pm 0.005 cm/gm^{1/3}$. Again only a subset of models are in agreement with these inversion results.

\begin{figure*}
\centering
\includegraphics[width=\hsize,clip]{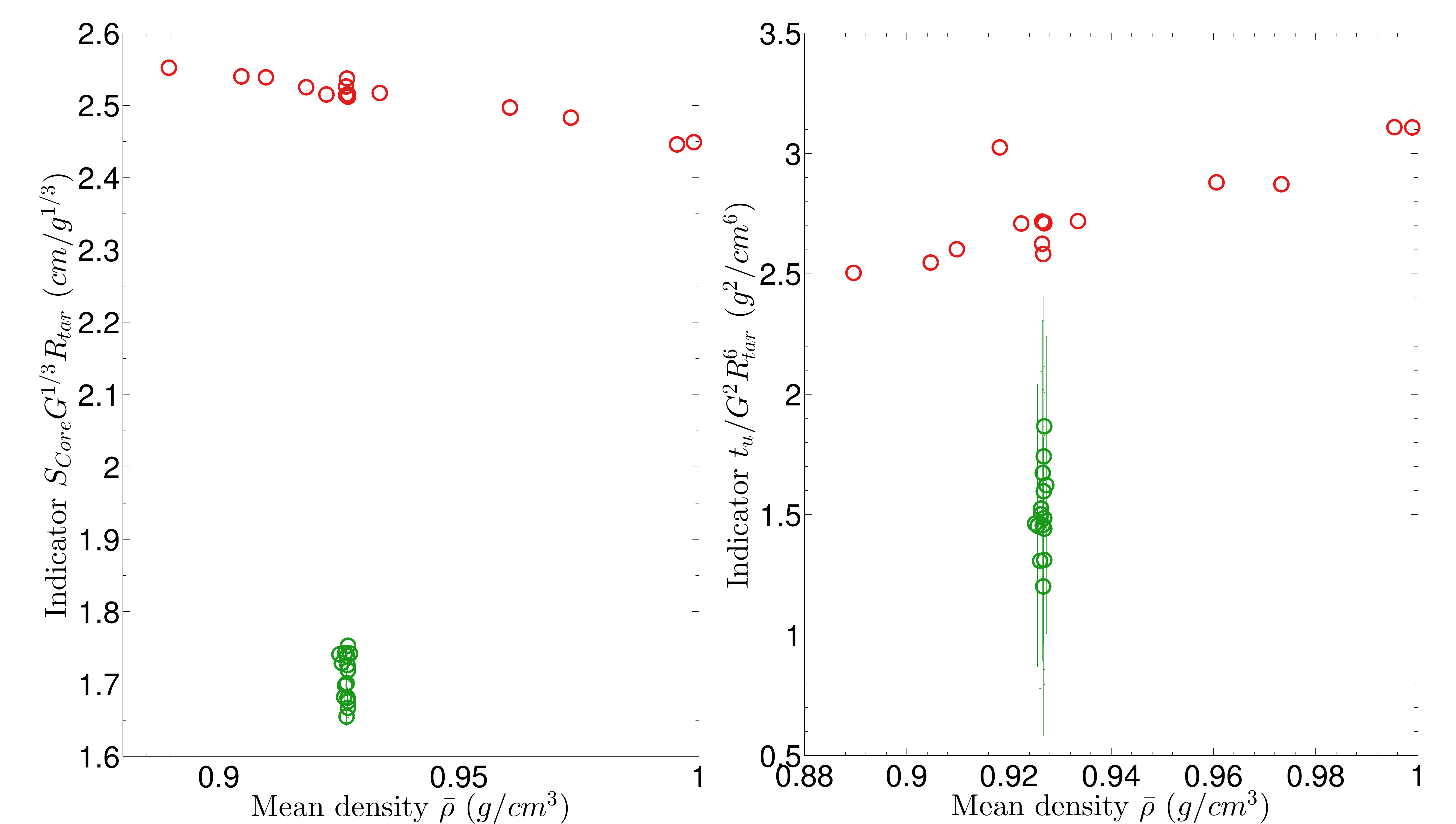}
\caption{Inversion results for Saxo. Left panel: $\bar{\rho}-S_{Core}G^{1/3}R_{tar}$ plane where the red circles illustrate the position of the reference models and the green crosses show the inversion results with their error bars. Right panel: $\bar{\rho}-t_{u}/(G^{2}R^{6}_{tar})$ plane following the same notations as the left panel.}
\label{FigResSaxo}       % Give a unique label
\end{figure*}

In fig. \ref{FigResSaxo}, we illustrate the inversion results for the $t_{u}$ and the $S_{Core}$ indicator. In this case, the results are stunningly different, since no model seems to agree with the inverted results, although the models were fitted to the observations with the same accuracy as for Doris. The difference is particularly significant for the $S_{Core}$ indicators and seems to indicate that something has been wrongly reproduced in the reference models. A small trend is seen with the mean density, as for the results for Doris, but the disagreement is still much larger. For the $t_{u}$ indicator, the value found is around $1.5\pm 1 g^{2}/cm^{6}$ while the $S_{Core}$ inversion leads to a result around $1.7 \pm 0.06 cm/g^{1/3}$. 
\section{Implication for the modelling of Doris and Saxo}
\label{sec-4}
\begin{figure}[h]
\centering
\includegraphics[width=\hsize,clip]{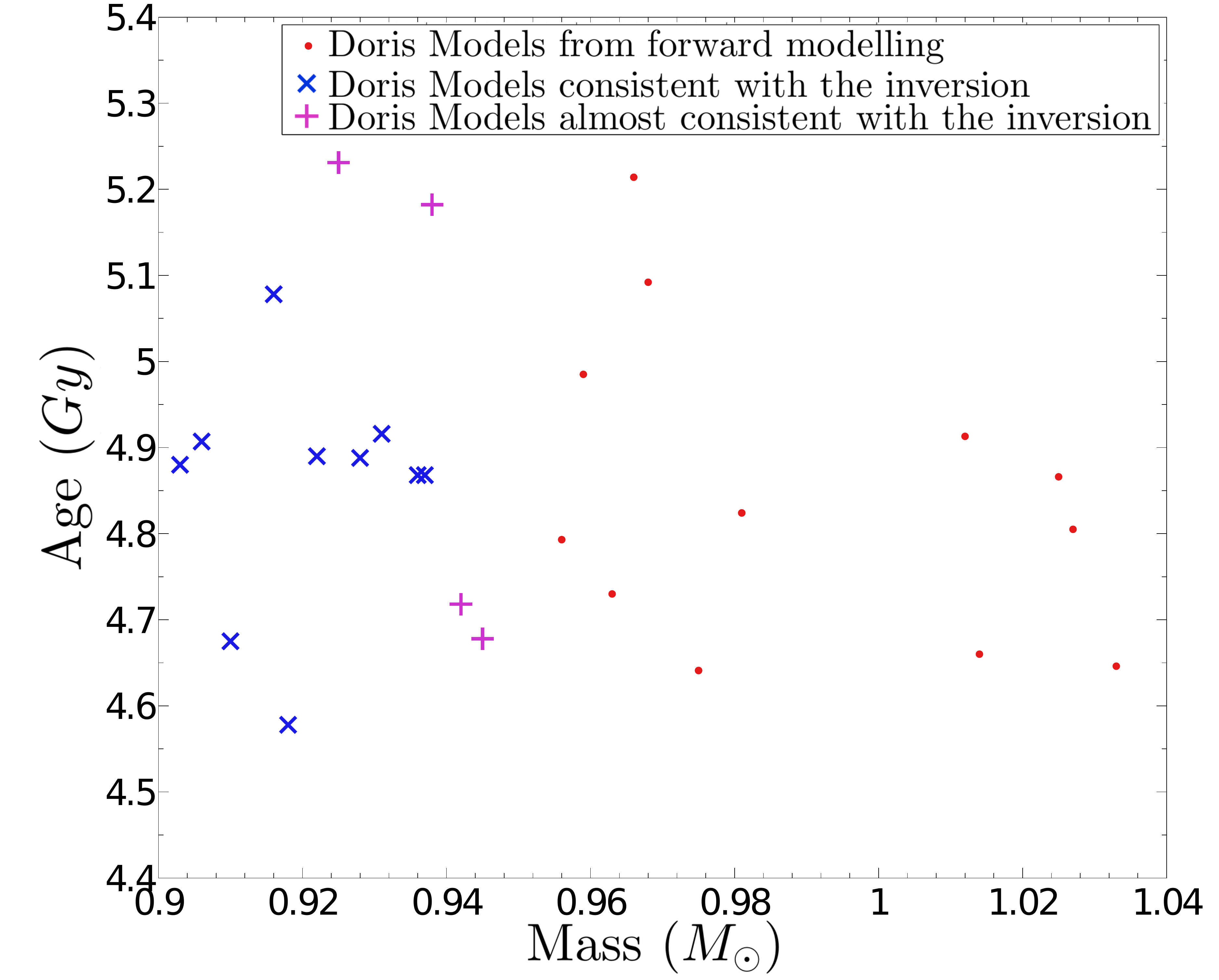}
\caption{Illustration in terms of mass and age of the models validated by the inversions for Doris. The blue crosses are the models in agreement with the inversions, the purple crosses are at the borders of the intervals validated by the inversions and the red dots are rejected by the inversions.}
\label{FigDispDoris}       % Give a unique label
\end{figure}

Using the inversion results for Doris, we could define a subsample of reference models, selected in agreement with the determinations of the mean density, the $t_{u}$ and the $S_{Core}$ indicators. These models are represented in Fig. \ref{FigDispDoris}, where the blue crosses represent the models of our sample which are in agreement with the inversion procedure, the purple crosses are models found in agreement with some but not all inversion results and the red dots are the models which disagree with the inversion results. We point out that all of these models included microscopic diffusion, so that the selection process is somewhat less spectacular than what is found in our study of $16$Cyg. It should be noted however, that models without microscopic diffusion for Doris are systematically rejected by both inversions and have thus not been included in this study. Another point worth mentioning is that unlike $16$Cyg, Doris has no determination of its surface helium abundance and thus further constraints could be brought on this model by determining the helium abundance of the convective envelope of this target. This could lead to further studies of this star and constrain the mixing inside this star, as is currently being done for $16$Cyg.

However, the use of constraints from the inversions leads us to select a subsample of models with a $2.5\%$ uncertainty in mass and $4.5\%$ uncertainty in age, illustrating the diagnostic process of the inversion technique and its efficiency in constraining fundamental parameters. The selection effect in age is very small since we did not plot models without microscopic diffusion, which were systematically rejected by the inversion technique. These uncertainties are of course internal error bars and do not take into account possible inaccuracies in the physical ingredients of the model. 

In the case of Saxo, no model seemed to fit the inverted constraints and thus a solution still has to be found. Moreover, the reliability of the frequencies has to be assessed before further inversions can be carried out. It is well known that the SOLA method is sensitive to outliers and it is also clear that any misfits to the frequencies would affect significantly the inversion results, and thus the diagnostic provided by the method. If these results are confirmed, then it would seem that something is clearly missing in the description of this target and further investigations would be needed. 

\section{Conclusions}
\label{sec-con}
In this paper, we applied the inversions of integrated quantities as defined in \cite{Reese} and \cite{Buldgentu} to two targets of the Kepler Legacy sample. This sample of $66$ dwarfs is considered to be the best sample of solar-like and F-type stars observed by Kepler. This means that these targets can be used as benchmark stars to constrain stellar physics and seismic modelling. Due to the high quality of the seismic data, they are very well suited for structural inversion techniques. We demonstrated this by applying our methods to Doris and Saxo and showed that there was indeed a diagnostic from the inversion technique. 

For Doris, this diagnostic allowed us to further constrain the fundamental parameters of this star and lead to a final uncertainty of $2.5\%$ in mass and $4.5\%$ in age. These results are, of course, model-dependent and to be considered as internal error bars, not taking into account potential errors in the physical ingredients of the stellar models. 

In the case of Saxo, a clear disagreement with the models was witnessed. The main problem with this result, but also with the seismic diagnostic for Doris, is linked to the uncertainties in the frequency determinations of the Kepler Legacy (see \cite{RoxCyg}). These results have then to be checked again once this problem will have been clarified.

\bibliography{GaelBuldgen}

\end{document}